\documentclass[10pt,aps,prl,twocolumn,groupedaddress,noshowkeys]{revtex4-1}
\usepackage{graphicx}
\usepackage{dcolumn}
\usepackage{bm}
\usepackage{amsmath}
\usepackage{hyperref}

\numberwithin{figure}{section}

\makeatletter 
    \renewcommand{\thefigure}{\@arabic\c@figure} 
\makeatother 

\newcolumntype{.}{D{.}{.}{-1}}

\begin{document}


\title{Quantum Simulation of Antiferromagnetic Spin Chains in an Optical Lattice}


\author{Jonathan Simon}
\author{Waseem S. Bakr}
\author{Ruichao Ma}
\author{M. Eric Tai}
\author{Philipp M. Preiss}
\author{Markus Greiner}
\email[]{greiner@physics.harvard.edu}
\affiliation{Department of Physics, Harvard University, Cambridge,
Massachusetts, 02138, USA}


\date{\today}

\begin{abstract}
Understanding exotic forms of magnetism in quantum mechanical systems is a
central goal of modern condensed matter physics, with implications from high
temperature superconductors to spintronic devices. Simulating magnetic
materials in the vicinity of a quantum phase transition is computationally
intractable on classical computers due to the extreme complexity arising from
quantum entanglement between the constituent magnetic spins. Here we employ a
degenerate Bose gas confined in an optical lattice to simulate a chain of
interacting \textit{quantum} Ising spins as they undergo a phase transition.
Strong spin interactions are achieved through a site-occupation to pseudo-spin
mapping. As we vary an applied field, quantum fluctuations drive a phase
transition from a paramagnetic phase into an antiferromagnetic phase. In the
paramagnetic phase the interaction between the spins is overwhelmed by the
applied field which aligns the spins. In the antiferromagnetic phase the
interaction dominates and produces staggered magnetic ordering. Magnetic
domain formation is observed through both in-situ site-resolved imaging and
noise correlation measurements. By demonstrating a route to quantum magnetism
in an optical lattice, this work should facilitate further investigations of
magnetic models using ultracold atoms, improving our understanding of real
magnetic materials.
\end{abstract}


\maketitle


Ensembles of quantum spins arranged on a lattice and coupled to one another
through magnetic interactions constitute a paradigmatic model-system in
condensed matter physics. Such systems produce a rich array of
magnetically-ordered ground states such as paramagnets, ferromagnets and
antiferromagnets. Certain geometries and interactions induce competition
between these orderings in the form of frustration, resulting in spin
liquids\cite{balents} and spin glasses\cite{binder_spin_glasses}, as well as
phases with topological order\cite{kitaev_anyons}. Varying system parameters
can induce quantum phase transitions between the various
phases\cite{sachdev_qpt}. A deeper understanding of the competition and
resulting transitions between magnetic phases would provide valuable insights
into the properties of complex materials such as high-temperature
superconductors\cite{anderson_resonating_valence_bond}, and more generally
into the intricate behaviours that can emerge when many simple quantum
mechanical objects interact with one another.

Studying quantum phase transitions of magnetic condensed matter systems is
hindered by the complex structure and interactions present in such systems, as
well as the difficulty of controllably varying system parameters. With a few
notable exceptions\cite{ruegg,coldea_criticality}, these issues make it
difficult to capture the physics of such systems with simple models.
Accordingly, there is a growing effort underway to realize condensed matter
simulators using cold atom systems\cite{lewenstein,bloch_rev} which are
understood from first principles. The exquisite control afforded by cold atom
experiments permits adiabatic tuning of such systems through quantum phase
transitions\cite{greiner_SF_MI,bloch_rev}, enabling investigations of
criticality\cite{sachdev_magnetism,chin_quantum_criticality} and
scaling\cite{chin_scale_invariance}. Time-resolved local
readout\cite{bakr_single_site,bakr_SF_MI,bloch_single_site} and
manipulation\cite{bloch_spin_addressing} provide direct access to local
dynamics and correlations. With this powerful toolbox in hand, considerable
attention has turned to understanding magnetic phase transitions using cold
atom quantum simulations.

Initial experimental efforts to observe quantum magnetism have focused on bulk
itinerant systems of ultracold
fermions\cite{ketterle_itinerant_ferromagnetism} and small, highly connected
spin-networks simulated with ion chains\cite{monroe_ising_chain}. Polar
molecules\cite{ni_polar_molecules} and Rydberg atoms\cite{pfau_rydberg} have
been the subject of preliminary investigations both experimentally and
theoretically\cite{zoller_atomic_ensembles,zoller_polar_molecules,zoller_rydberg_simulator}
as alternatives to ground-state atoms with stronger, longer-range
interactions. There has also been initial success in detecting ordered states
which are artificially prepared through patterned
loading\cite{lee,foelling_strongly_correlated,sengstock_hexagonal,ketterle_itinerant_ferromagnetism}
and double-well\cite{foelling_second_order} experiments.

In this work, we simulate a 1D chain of interacting Ising spins by mapping
doublon-hole excitations of a Mott
insulator\cite{fisher_SF_MI,jaksch,greiner_SF_MI} of spinless bosons in a
tilted 1D optical lattice\cite{sachdev_MI_electric_fields} onto a pseudo-spin degree of freedom. This is in
contrast to the commonly considered approach in which the magnetic spins are
represented by two internal states of the atoms, and nearest-neighbor
spin-spin interactions result from super-exchange
couplings\cite{duan_spin_exchange}. Super-exchange
interactions in cold atoms are quite weak, though they have been successfully
observed in double well systems\cite{foelling_superexchange}. The approach presented here has the benefit
of a dynamical timescale set by the tunneling rate $t$, rather than the
super-exchange interaction $t^2/U$, where $U$ is the onsite interaction energy.
Combining the faster dynamics with the high spatial resolution afforded by a
quantum gas microscope\cite{bakr_single_site}, we are able to directly observe
transitions between paramagnetic and antiferromagnetic phases as spin-spin
interactions compete with applied fields.

One of the primary concerns in studying transitions to magnetic states in cold
atomic gases is the apparent difficulty of reaching the requisite
temperatures\cite{capogrosso_critical_entropy}. These spin temperatures are at
the edge of experimental
reach\cite{ketterle_spin_gradient,ketterle_demag_cooling}, and further cooling
of lattice-spins remains an active field of research\cite{mckay_cooling}.
However, ultracold gases in optical lattices are effectively isolated from
their environment, and as such it is entropy and not temperature which is
constant as system parameters are tuned. Spin-polarized Mott insulators have
been demonstrated with defect densities approaching the percent
level\cite{bakr_SF_MI,bloch_single_site}, corresponding to configurational
entropy far below the spin entropy required for magnetic ordering (see SI).
This allows us to use such a Mott insulator to initialize a magnetic system
with low spin entropy.  We engineer a magnetic Hamiltonian whose paramagnetic
ground state possesses good overlap with the initial Mott state, and
subsequently tune it through a quantum phase
transition\cite{cirac_spin_hamiltonians} to produce an antiferromagnetic
state. The difficulty of cooling lattice spins is thus replaced with the
necessity of performing sufficiently slow adiabatic ramps to minimize diabatic
crossings of manybody energy gaps.

\begin{figure}
    \centering
    \includegraphics[width=7cm]{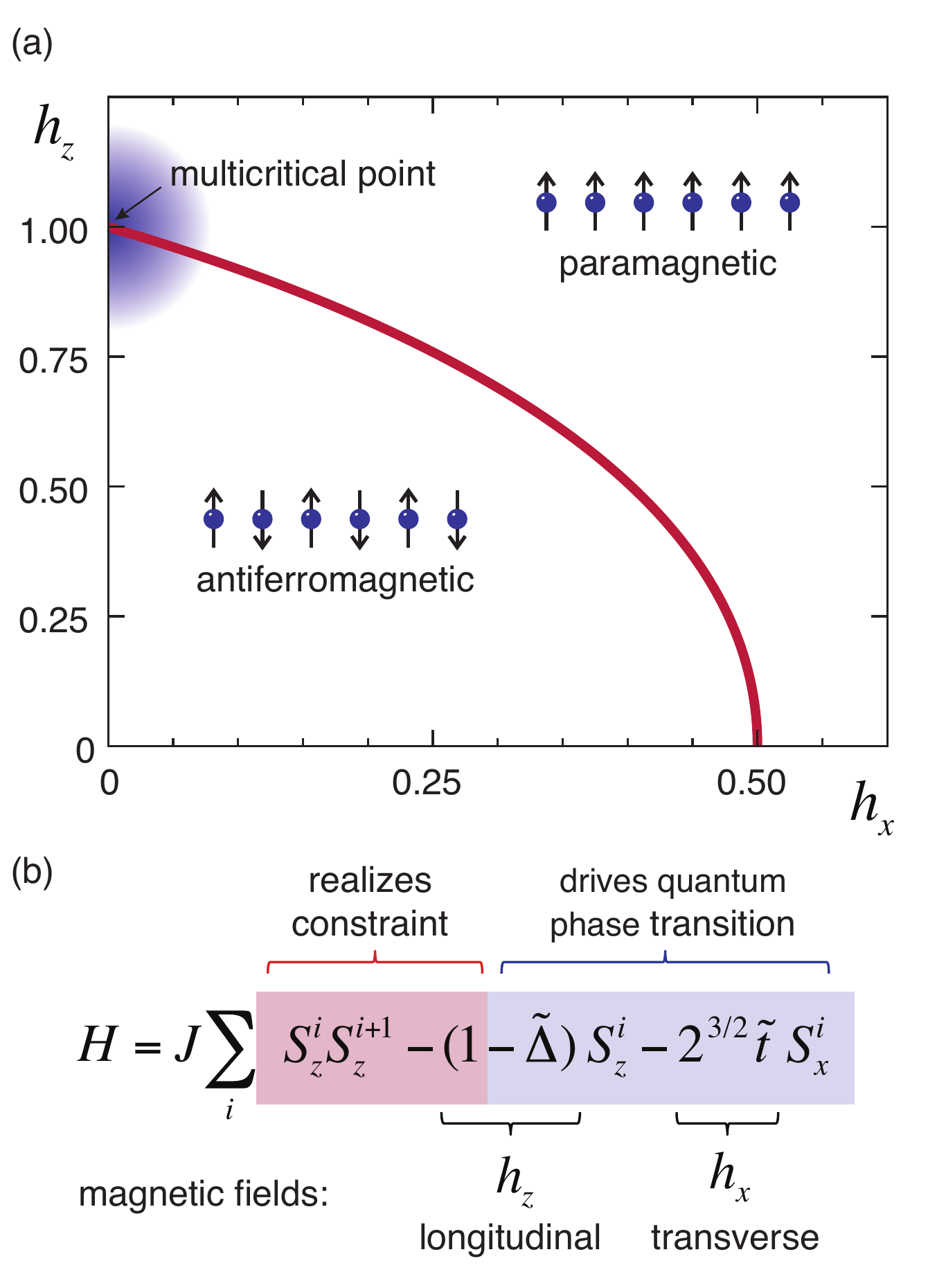}
    \caption{\label{fig1} \textbf{Spin model and its phase diagram.}
    \textbf{a}. An antiferromagnetic 1D Ising chain in longitudinal ($h_z$)
    and transverse ($h_x$) magnetic fields exhibits two phases at zero
    temperature\cite{ovchinnikov,landau}. For weak applied fields the
    interactions between the spins drive the system to form an
    antiferromagnet. For strong applied fields the spins align to the field
    and produce a paramagnet. These two phases are separated by a second order
    phase transition (red line), except at the multicritical point
    $(h_z,h_x)=(1,0)$, where the lack of a transverse field makes the
    transition classical and first order. The region that can be accessed in
    our experiment, in the vicinity of $(h_z, h_x) = (1,0)$, is highlighted in
    blue. \textbf{b}. In this neighborhood the Hamiltonian may be decomposed
    into a constraint term that prevents adjacent spin-flips (red highlight),
    and fields that drive the phase transition (blue highlight).}
\end{figure}

\subsection{Ising Interactions in a Tilted Optical Lattice}
The quantum Ising model is a paradigmatic model of magnetism, and an Ising
chain is one of the simplest many-body systems to exhibit a quantum phase
transition. The Hamiltonian describing a 1D antiferromagnetic Ising chain in
the presence of an applied magnetic field is given by:
\[H = J \sum_i \left(S_z^i S_z^{i+1} - h_z S_z^i - h_x S_x^i\right)\]
Here $S_z^i$ ($S_x^i$) is the $z$ ($x$) spin-projection operator at site $i$,
and $h_z$ ($h_x$) is the $z$ ($x$)-component of the magnetic field applied to
site $i$. The zero temperature phase diagram of the
model\cite{landau,ovchinnikov} is shown in Fig.~\ref{fig1}a, for a homogeneous
applied field $(h_z,h_x)$. For small applied fields, the magnetic interactions
induce staggered ordering of the spins, producing an antiferromagnet (AF). For
large applied fields, the field overwhelms the interactions, and all spins
align to the field, producing a paramagnet (PM).

Our approach to constructing a magnetic Hamiltonian was proposed by Sachdev
\textit{et al.}\cite{sachdev_MI_electric_fields}, in the context of
experiments by Greiner \textit{et al.}\cite{greiner_SF_MI}, where a gradient was
applied to measure the insulating properties of the Mott state. Sachdev
\textit{et al.} showed that under the influence of such field gradients, the dynamics of a 1D
Mott insulator map onto the aforementioned Ising model (see Methods) in the
neighborhood of $(h_z, h_x)=(1,0)$ (Fig.~\ref{fig1}b).

\begin{figure*}
    \centering
    \includegraphics[width=15cm]{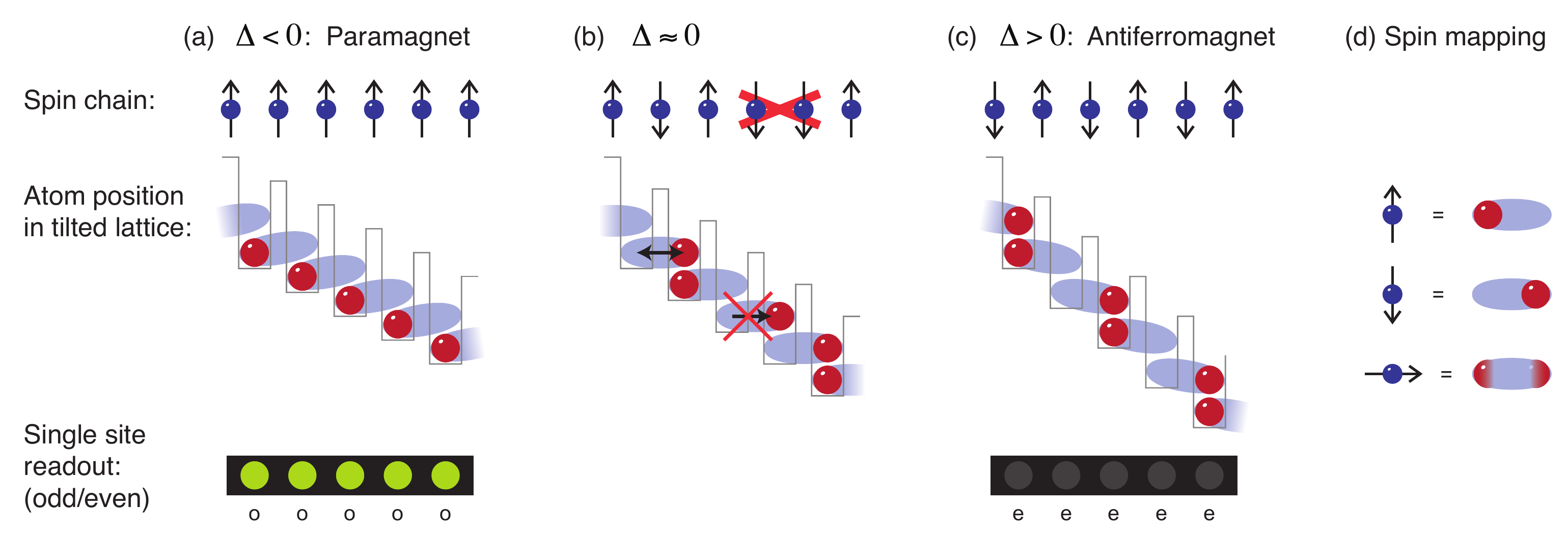}
    \caption{\label{fig2} \textbf{Tilted Hubbard model and mapping to spin
    model.} \textbf{a}. Middle row: When a Mott insulator is placed in a
    tilted lattice, it remains in a state with one atom per lattice site until
    the tilt per site $E$ reaches the onsite interaction energy $U$.
    \textbf{b}. At this point the energy cost $\Delta=E-U$ to move to the
    neighboring site vanishes, and the atoms begin to tunnel resonantly to
    reduce their energy. An atom, however, can only tunnel to a neighboring
    site if the atom on that site has not itself tunneled away. If no atom is
    present on the neighboring site, the tunneling process is suppressed by
    the energy gap $U$. This creates a strong constraint and leads to the
    formation of entangled states. \textbf{c}. As the tilt is increased
    further, the system transitions into a doubly degenerate staggered phase.
    \textbf{d}. This system may be mapped onto a model of interacting spin-1/2
    particles, where the two spin states correspond to the two possible
    positions of each atom. In the spin model, the aforementioned constraint
    forbids adjacent down spins, realizing a spin-spin interaction. The
    initial Mott insulator now corresponds to a paramagnetic phase with all
    spins aligned upwards to a large magnetic field (see top row), the state
    at resonant tilt corresponds to a non-trivial (critical) spin
    configuration, and staggered ordering at even larger tilt corresponds to
    an anti-ferromagnetic phase. Bottom row: The phases can be detected by
    single lattice site imaging. Because the imaging system is sensitive only
    to the parity of the atom number, paramagnetic domains appear bright
    \textbf{a}., and anti-ferromagnetic domains appear dark \textbf{c}.}
\end{figure*}

In the Mott insulator regime ($U \gg t$) it is energetically forbidden for the
atoms to tunnel as long as the tilt per lattice site, $E$, differs from the
onsite atom-atom interaction $U$. Hence, the system remains in a state with
one atom per lattice site for $E<U$ (Fig.~\ref{fig2}a). As the tilt approaches
the interaction strength ($E=U$), each atom is free to tunnel onto its
neighbor, so long as its neighbor has not itself tunneled (Fig.~\ref{fig2}b).
This nearest-neighbor constraint is the source of the effective spin-spin
interaction. If the tilt $E$ is increased sufficiently slowly through the
transition so as to keep the system near the many-body ground state, density
wave ordering results (Fig.~\ref{fig2}c).

The mapping onto a spin-1/2 model arises as each atom has only two possible
positions: an atom that has not tunneled corresponds to an ``up'' spin, and an
atom that has tunneled corresponds to a ``down'' spin. Fig.~\ref{fig2}d shows
the spin configurations that correspond to various atom distributions in the
optical lattice. The transition from a uniform phase at small tilt to a
density wave phase at large tilt then corresponds to a transition from a
paramagnetic phase to an antiferromagnetic phase in the spin model. The
longitudinal field $h_z$ thus arises from the lattice tilt, and the transverse
field $h_x$ from tunneling. As derived in the Methods, the mapping between
Bose-Hubbard and spin models is given by $(h_z, h_x) = (1 - \tilde \Delta,
2^{3/2} \tilde t)$, $\tilde t = t/J$, $\tilde \Delta = \Delta/J = (E-U)/J$,
with $t$ the single-particle tunneling rate, and $J\sim U$ the constraint term.

Spatially varying tilts in the optical lattice can give rise to site-to-site
variation of $h_z$. Such inhomogeneity would impact the critical behaviour by
breaking the translational symmetry, inducing different sites to transition at
different applied tilts. Accordingly, the resulting many-body energy gaps,
dynamical timescales\cite{ma,dziarmaga}, and entropy of
entanglement\cite{kitaev_entanglement} would differ from the homogeneous case.
Controlling such inhomogeneities is thus crucial for studies of magnetism.

The mapping of the Hubbard model onto the spin-model breaks down away from
$(h_z, h_x) \sim (1,0)$. This is because states with three atoms on a lattice
site are not within the Hilbert space that maps to the spin model. As such we
can study ground state dynamics and low-energy excitations of the equivalent
Ising model, but not high-energy excitations associated with adjacent flipped
spins. These constraints admit investigation of the Ising physics only in the
neighborhood of the multicritical point $(h_z, h_x) \sim (1,0)$. This regime
is of particular theoretical interest as the model is here not exactly
solvable\cite{ovchinnikov}. It is nonetheless in the Ising universality
class\cite{sachdev_MI_electric_fields}, and so a study of its critical physics
would provide insight into the behaviours of the more commonly considered
transverse ($h_z=0$) Ising model.

\subsection{Extracting Spin Observables}
We locally detect magnetic ordering by utilizing our quantum gas microscope,
capable of resolving individual lattice sites. The microscope is sensitive
only to the parity of the site occupation number\cite{bakr_single_site}, and
so PM domains (with one atom per lattice site) should appear as entirely
bright regions, and AF domains (with alternating 0-2-0-2 occupation) as
entirely dark regions. In the spin language, the detection parity operator
$P^i$ measures the spin-spin correlation between adjacent spins:
\[P^i = 4 S_z^{i-1} S_z^i\]
Throughout this article we will characterize our spin-ordering primarily via
the probability that site i has odd occupation $p_{\text{odd}}^i = \frac12 (1
+ \langle P^i \rangle)$. Taking the spin constraint into account, the chain
average of $p_{\text{odd}}^i$ is equivalent to a chain-averaged measurement of
$\langle S_z^i \rangle$, the mean $z$-projection of the spin:
$\overline{\langle S_z^i \rangle} = \frac12 \overline{p_{\text{odd}}^i}$ .
Here angle brackets denote realization-averages and bars denote
chain-averages.

This parity measurement allows us to locally identify magnetic domains and
estimate their size. This, however, is not a direct measurement of the AF
order parameter described in Ref. \cite{sachdev_MI_electric_fields}, and does
not reflect the broken symmetry in the AF phase (see Methods). Accordingly, we
also study the AF order parameter more directly through 1D quantum noise
interferometry\cite{lukin_noise_correlation}.

\subsection{Observing the Phase Transition}
\begin{figure}
    \centering
    \includegraphics[width=8.5cm]{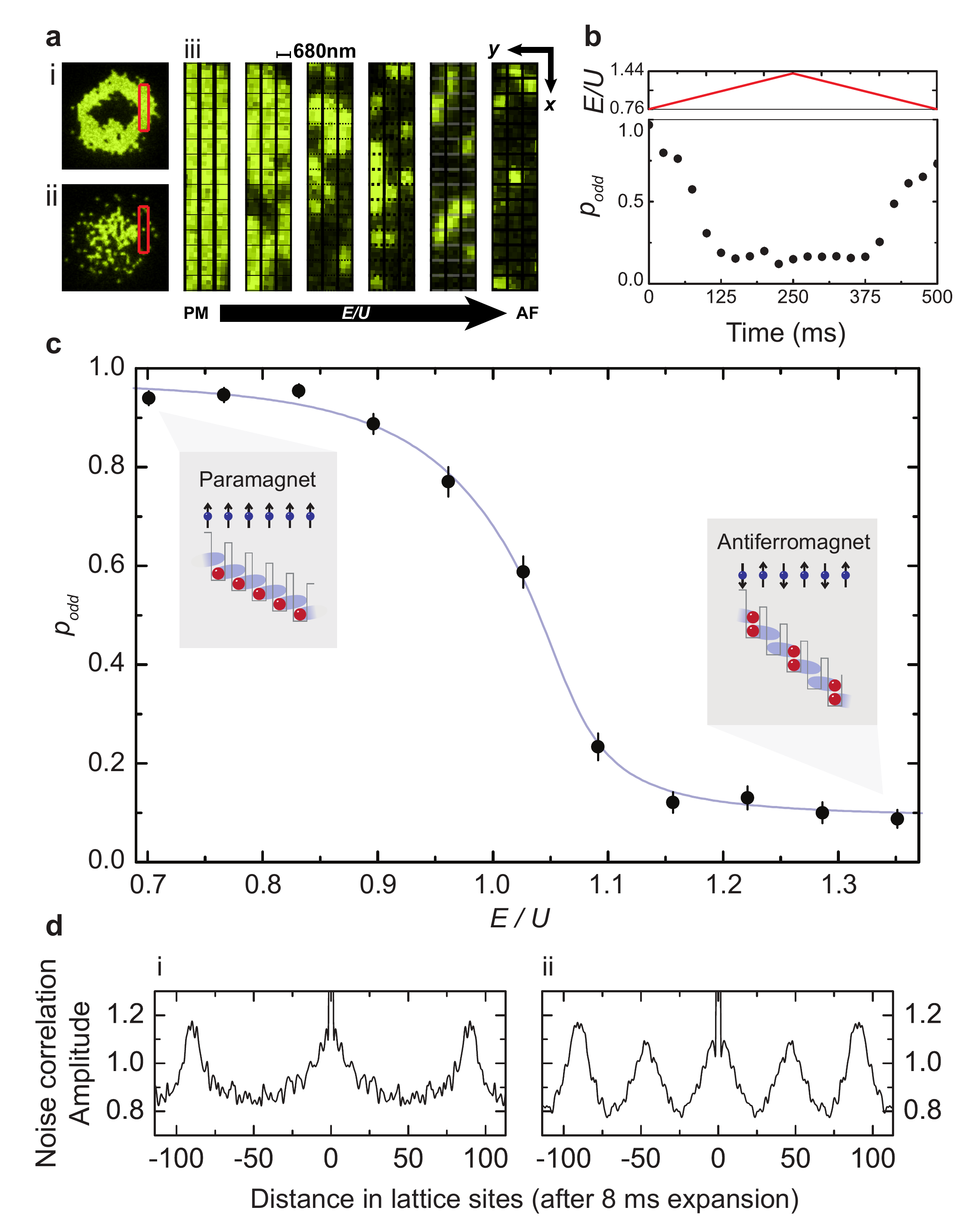}
    \caption{\label{fig3} \textbf{Probing the paramagnet to antiferromagnet
    phase transition.} \textbf{a}. Representative single-shot images as the
    tilt is swept adiabatically through the phase transition in 250 ms. The
    upper image (i) is near-perfect $n=1$ (bright) and $n=2$ (center dark)
    Mott insulator shells in PM phase. The lower image (ii) is the
    ``inverted'' shell structure characteristic of the staggered ordering of
    the AF phase after a tilt along the $x$-direction. The inversion occurs
    because chains of sites in a shell with $N$ atoms per site are converted
    into a staggered phase wherein sites alternate between $N-1$ and $N+1$
    atoms, and so a shell with even occupation becomes a region of odd
    occupation, and vice-versa. The remaining pictures (iii) are several
    chains (within the red rectangles in i and ii) of the $N=1$ shell at
    various points during the sweep, $t= 0$, 50, 100, 150, 175 and 250 ms,
    showing AF domain formation. \textbf{b}. To demonstrate the reversibility
    of the transition, we adiabatically ramp from the PM phase into the AF
    phase and back in 500 ms. The probability that a site in the $N=1$ shell
    has odd occupation at various points during the ramp is observed to drop,
    and subsequently revive, as expected when the system leaves and then
    returns to the PM phase. \textbf{c}. A closer look at the PM to AF quantum
    phase transition within an $N=1$ shell, showing $p_{\text{odd}}$ vs. tilt. Errorbars
    reflect $1\sigma$ statistical errors in the region-averaged mean
    $p_{\text{odd}}$. The blue curve is a guide for the eye. \textbf{d}. Noise
    correlation measurement after 8 ms time of flight expansion along the
    chains. (i) In the PM phase, peaks at momentum $h/a$ correspond to a
    periodicity of one lattice site before expansion, characteristic of a Mott
    insulator\cite{foelling_noise_interferometry}. (ii) In the AF phase,
    additional peaks at momentum $h/2a$ indicate the existence of staggered
    ordering, with a periodicity of two lattice sites.}
\end{figure}

Our experiments begin with a Mott insulator of $^{87}$Rb atoms in a two
dimensional optical lattice with spacing $a=680\,\text{nm}$ and a depth of
$35E_r$, in the focal plane of a high resolution imaging system which allows
detection of single atoms on individual lattice sites as described in previous
work\cite{bakr_single_site,bakr_SF_MI}. The lattice recoil energy is given by
$E_r=h^2/8ma^2$, where $h$ is Planck's constant and $m$ the mass of $^{87}$Rb.
We generate our effective $h_z$ by tilting the lattice potential by $E$ per
lattice-site, which is achieved via a magnetic field gradient along the
$x$-direction (defined in Fig.~\ref{fig3}a). This gradient is applied in two
steps—first a fast ramp (in 8 ms) to just below the transition
point\cite{sachdev_MI_electric_fields,ovchinnikov} at $E=U+1.85 t$
($h_z=1-0.66 h_x$), followed by a slow linear ramp (in 250 ms) across the
transition. Before starting the slow gradient ramp, the lattice depth along
the $y$-direction is increased (in 2 ms) typically to $45(7)E_r$, while the
depth along the $x$-direction is reduced to $14(1)E_r$. This decouples the
system into 1D chains with significant tunneling only along their lengths.
Simultaneously, we compensate the tilt inhomogeneity arising from harmonic
confinement, leaving only residual inhomogeneity arising from our lattice
projection method\cite{bakr_single_site}. We then probe the system by stopping
the ramp at various points and observing spin ordering, first of the entire
cloud, and then specializing to the transitions of individual lattice sites in
a particular chain.

We initiate the gradient ramp on the paramagnetic side of the phase transition
(typically at $E/U=0.7$), as the initial Mott state has good overlap with the
paramagnetic ground state (Fig.~\ref{fig3}ai). At the end of the ramp
($E/U=1.2$), we observe an even occupation with probability 0.90(2)
(Fig.~\ref{fig3}aii), as expected for an AF phase in the magnetic model where
the spin-spin interaction overwhelms the effective field $h_z$. In between,
density-wave (AF) ordered regions begin to form, as shown in
Fig.~\ref{fig3}aiii. Fig.~\ref{fig3}c shows $p_{\text{odd}}$ at various times
during this ramp. A crucial characteristic of an adiabatic transition is that
it is \textit{reversible}. Fig. 3b shows $p_{\text{odd}}$ during a ramp from a PM to an
AF and back. The recovery of the singly occupied sites is evidence of the
reversibility of the process, and hence that the state at the end of the
forward ramp is in fact an antiferromagnet.

We directly verify the existence of staggered ordering in the AF phase via a
1D quantum noise correlation measurement\cite{lukin_noise_correlation}. We
perform this measurement by increasing the lattice depth along the chains to
$35E_r$ within 5 ms and then rapidly switching off that lattice to realize a
1D expansion. The resulting spatial autocorrelation is plotted in
Fig.~\ref{fig3}d at both the beginning (i) and end (ii) of the ramp from the
PM phase to the AF phase. In the PM phase the spectrum exhibits peaks at
momentum difference $p=h/a$, characteristic of a Mott
insulator\cite{foelling_noise_interferometry}. In the AF phase peaks at $p=h/2a$
appear, indicative of the emergence of a spatial ordering with twice the
wavelength. In principle the mean domain size can be extracted from the
$p=h/2a$ peak width, however our measurement is broadened by both finite
expansion time and aberration arising from the fact that the 1D expansion is
performed not in free space but in slightly corrugated confining tubes.

\subsection{Single-Site Study of the Transition}
\begin{figure}
    \centering
    \includegraphics[width=8.5cm]{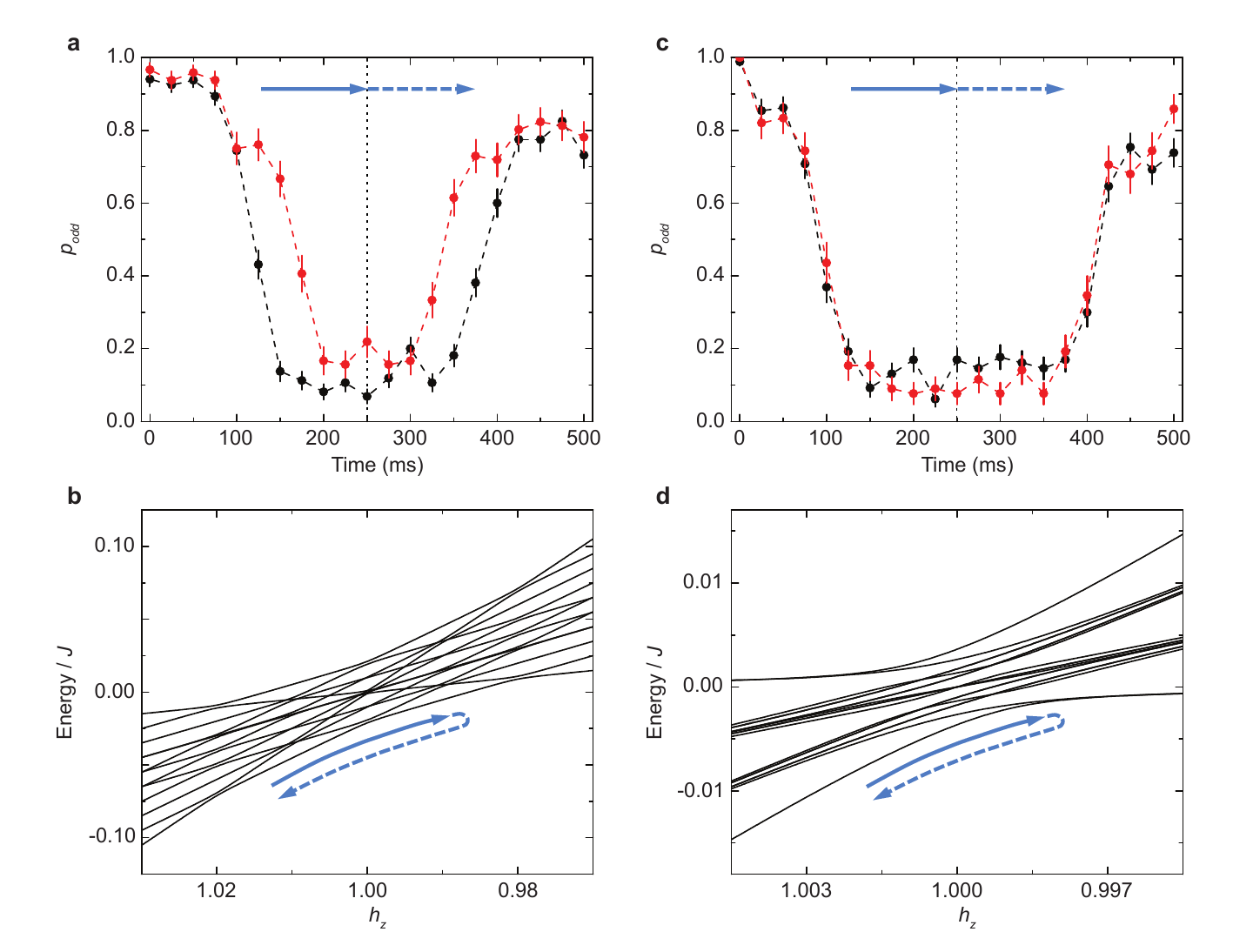}
    \caption{\label{fig4} \textbf{Impact of harmonic confinement.} For a ramp
    (see Fig.~\ref{fig3}b) across the transition (in 250 ms) and back,
    \textbf{a}. harmonic confinement broadens the transition, inducing rows of
    the cloud seven lattice sites apart to undergo the transition at different
    applied tilts. \textbf{b}. Corresponding energy spectrum of a 1D Ising
    chain in a longitudinal field gradient of 0.01 per lattice site,
    reflecting seven spins and open boundary conditions. Each avoided crossing
    of the lowest energy state corresponds to a single spin-flip with energy
    gap $h_x=0.001$. \textbf{c}. Once the confinement has been properly
    compensated, the average transition curves from the two rows overlap. Not
    apparent from these averaged curves is a small amount of residual tilt
    inhomogeneity. \textbf{d}. Energy spectrum for a homogeneous 1D Ising
    chain of six spins, with periodic boundary conditions. In contrast to
    \textbf{b}., the single avoided crossing drives all spin-flips
    simultaneously, with a gap that decreases with increasing system size, as
    expected for the critical slowdown near a quantum phase transition. All
    errorbars are $1\sigma$ statistical uncertainties derived from the mean of
    $p_{\text{odd}}$ averaged over a region.}
\end{figure}

A high resolution study reveals that in the presence of harmonic confinement the spins
undergo the transition sequentially due to the resulting spatial variation of
the effective longitudinal field. Fig.~\ref{fig4}a shows $p_{\text{odd}}$ vs.
tilt for two different rows of a harmonically confined Mott insulator,
separated by seven lattice sites. These two rows tune through resonance at
different tilts, as can be understood from the energy level diagram
Fig.~\ref{fig4}b. To realize a homogeneous field Ising model we eliminate the
harmonic confinement (see Methods) immediately before the slow ramp into the
AF phase. The homogeneity is now only limited by residual lattice beam
disorder, and accordingly, Fig.~\ref{fig4}c demonstrates that different rows
transition almost simultaneously, as anticipated theoretically
(Fig.~\ref{fig4}d).

\begin{figure}
    \centering
    \includegraphics[width=8.5cm]{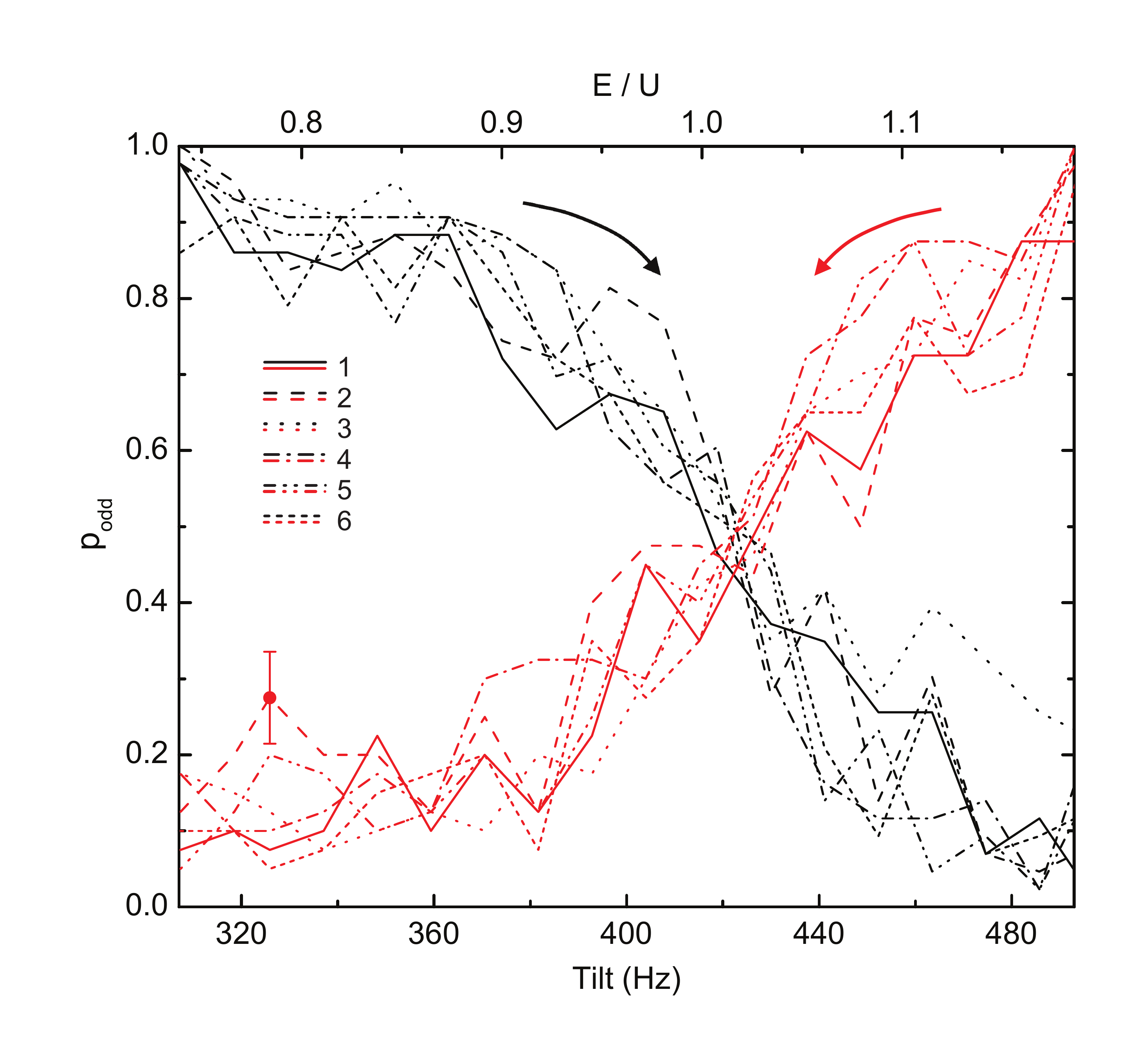}
    \caption{\label{fig5} \textbf{Site-resolved transition in near-homogeneous
    Ising model.} The probability of odd occupation of six individual lattice
    sites forming a contiguous 1D chain, versus the tilt, shown for both
    forward (black) and reverse (red) ramps. The spins transition at the same
    applied field to within the curve width, set by quantum fluctuations. A
    typical $1\sigma$ statistical errorbar is shown. The single-site widths
    are consistent with a longitudinal lattice depth of $14(1)E_r$, in
    agreement with a Kapitza-Dirac measurement of $15(2)E_r$. A large local
    potential step at a particular lattice would be reflected as a shift in
    either the forward or reverse $p_{\text{odd}}$ curve at that site. The
    reverse curve demonstrates our ability to adiabatically prepare the
    highest energy state of the restricted spin Hamiltonian: the system
    exhibits PM ordering on the AF side of the transition, and AF ordering on
    the PM side of the transition.}
\end{figure}

After compensating the harmonic confinement, we use high resolution imaging to
study the transition on the single-site level. This allows us to focus on a
single six-site chain with particularly low inhomogeneity, which we will study
for the remainder of this article. We identify such a chain by imaging
individual lattice sites as the system is tuned across the PM-AF transition.
Fig.~\ref{fig5} shows the average occupation of each of the six adjacent sites
(black curves), versus tilt, for a 250 ms ramp across the transition. The
r.m.s. variation in the fitted centers is 6 Hz, significantly less than their
mean 10\%--90\% width of 105(30) Hz, corresponding to the effective transverse
field $2^{3/2} t = 28\,\text{Hz}$. By quickly jumping across the transition
with tunneling inhibited, and then ramping slowly across the transition in
reverse with tunneling allowed (red curves, taken under slightly different
conditions), we are able to rule out large, localized potential steps that
would otherwise prevent individual spins from flipping. The curves in
Fig.~\ref{fig5} provide our best estimate of the inhomogeneity. However, exact
determination of the site-to-site disorder using this technique is complicated
by the many-body nature of the observed transition. New spectroscopic
techniques such as single-site modulation spectroscopy would need to be
developed to ensure that the inhomogeneities are small enough to study
criticality in long, homogeneous Ising chains.

\subsection{Domain Formation in a 1D Ising Chain}
\begin{figure}[!h]
    \centering
    \includegraphics[width=8cm]{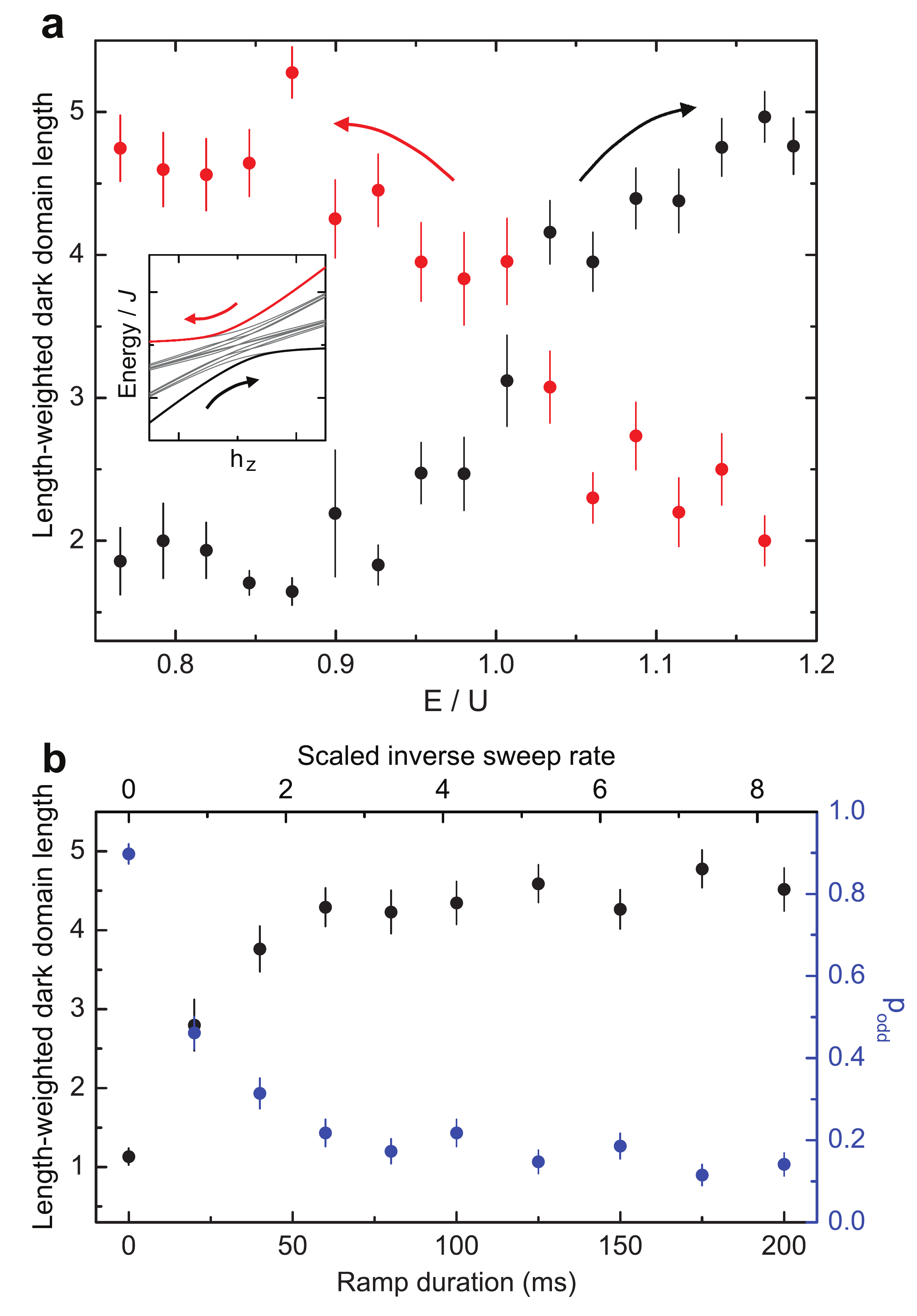}
    \caption{\label{fig6} \textbf{Dynamics of antiferromagnetic domain
    formation.} Within a single six-site chain with low disorder, \textbf{a}. shows the
    mean dark-domain length as a function of the tilt in units of $U$, for
    both forward (black) and reverse (red) ramps. As the system enters the AF
    phase the mean dark domain length grows until it approaches the chain
    length. Domain formation in the reverse ramp, beginning on the AF side of
    the transition, demonstrates the adiabatic generation of AF domains on the
    PM side of the transition, corresponding to the highest energy state of
    the spin Hamiltonian (inset). Within the same chain, \textbf{b}. shows
    $p_{\text{odd}}$ (blue) and dark domain length (black) versus the duration
    $T_{\text{ramp}}$ of the ramp from $E/U=0.7$ to $E/U=1.2$. The top axis
    shows the scaled inverse sweep rate $\alpha = 8\pi^2 t^2/(\Delta
    E/T_{\text{ramp}})$, where $\Delta E$ is the sweep range in Hz, and $t$ is
    the tunneling rate, in Hz, along the chain. The characteristic timescale
    for domain formation is $\alpha \approx 2$, or $T_{\text{ramp}} \approx 50
    \, \text{ms}$, indicating that tunneling along the chain is the source of
    the quantum fluctuations that drive domain formation. Errorbars for the
    dark chain lengths are $1\sigma$ statistical uncertainties, arising from
    the number of detected domains of each length. Those for $p_{\text{odd}}$
    are the $1\sigma$ statistical uncertainty in the mean of the six-site
    chain.}
\end{figure}

Quantum fluctuations induce the formation of AF domains as the homogeneous
chain is ramped through the transition. As discussed previously, these domains
will appear as uninterrupted strings of dark lattice sites. Fig.~\ref{fig6}a
shows the observed mean length-weighted dark domain length extracted from 43
single-shot images per tilt, as the system is ramped from the PM phase into
the AF phase. The dark domain length is here defined as the number of
contiguous dark sites (see SI). On the AF side of the transition, the mean
dark domain length grows to 4.9(2) sites, giving evidence that the average AF
domain size approaches the system size.

We next investigate the impact of ramp rate on the transition from the PM
phase into the AF phase in the homogeneous chain. The blue points in
Fig.~\ref{fig6}b show $p_{\text{odd}}$ as a function of ramp speed across the
transition, which may be understood qualitatively as the fraction of the
system that has not transitioned into AF domains of any size. The time
required to flip the spins is $\sim\!50\,\text{ms}$, consistent with tunneling
induced quantum fluctuations driving the transition. The black points in
Fig.~\ref{fig6}b are the mean length-weighted dark domain length as a function
of ramp rate. As above, the mean dark domain length saturates at 4.8(2), near
the system size of six sites.  The remaining defects likely result from
imperfect overlap of the initial MI with the PM state at finite $\tilde
\Delta$, as well as defects of the initial MI and heating during the ramp.

While the antiferromagnetic domain formation discussed thus far occurs in
spin-chains that remain in a quantum state near the many-body ground state, we
can also produce antiferromagnetic domains that correspond to the highest
energy state of the constrained Hilbert space. This is achieved by starting
with the Mott insulator and rapidly ramping the field gradient through the
transition point with tunneling inhibited, then adiabatically ramping back
with tunneling permitted. This prepares a PM on the AF side of the transition,
and adiabatically converts it into an AF on the PM side. The resulting data
are shown in red, in Fig.~\ref{fig5} and Fig.~\ref{fig6}a, demonstrating that
these high energy states are sufficiently long-lived to support domain
formation. Similar ideas have been proposed for preparation of
difficult-to-access many-body states using the highest energy states of
Hamiltonians with easily prepared ground states\cite{demler}.

\subsection{Conclusions and Outlook}
We have experimentally realized a quantum simulation of an Ising chain in the
presence of longitudinal and transverse fields. By varying the applied
longitudinal field, we drive a transition between PM and AF phases, and verify
the formation of spin domains via both direct \textit{in situ} imaging, and
noise-correlation in expansion. We study the adiabaticity requirements for
transition dynamics, and observe a timescale consistent with tunneling-induced
quantum fluctuations. By rapidly tuning through the transition and ramping
back across it slowly, we prepare the highest energy state of a many-body spin
Hamiltonian.

We introduce and implement a novel route to studying low entropy magnetism in
optical lattices. Complexities associated with cooling of spin
mixtures\cite{ketterle_spin_gradient,ketterle_demag_cooling,mckay_cooling} are
circumvented by employing a low entropy, gapped Mott insulator as an initially
spin-polarized state, and then adiabatically opening a spin degree of
freedom\cite{cirac_spin_hamiltonians}. This recipe is directly applicable to
more traditional approaches to quantum magnetism, including those employing
super-exchange interactions\cite{duan_spin_exchange}.

The spin-mapping demonstrated here opens a new avenue for future work on
quantum magnetism and control. Strong effective magnetic interactions make
possible the creation of states with coherently generated long-range order.
Combined with a lower-disorder lattice, in-depth studies of criticality become
possible. Tilting by the band excitation energy will enable studies of the
transverse Ising model\cite{wimberger}. It will be interesting to investigate
the impact of various types of controlled disorder on criticality and
transition dynamics, as well as the possible existence of non-thermalizing
states\cite{cirac_thermalization}. A particularly intriguing direction is the
extension of the tilted Mott insulator physics to higher dimensions. The
simple square lattice geometry will provide access to phases with longitudinal
density wave ordering and transverse
superfluidity\cite{sachdev_MI_electric_fields}. More sophisticated geometries
will produce frustrated systems with novel quantum liquid and dimer covered
ground states\cite{pielawa}.

\begin{acknowledgments}
We would like to thank E. Demler, T. Kitagawa, M. D. Lukin, S. Pielawa, and S.
Sachdev for stimulating discussions. This work was supported by grants from
the Army Research Office with funding from the DARPA OLE program, an AFOSR
MURI program, and by grants from the NSF.
\end{acknowledgments}

\bibliography{mag_paper_arXiv}

\appendix
\section{Methods}
\makeatletter 
    \renewcommand{\thefigure}{S\@arabic\c@figure} 
\makeatother 
\subsection{Mapping onto the Spin Model}
We follow Sachdev \textit{et al.} in formally mapping a 1D Mott insulator of
spinless bosons in a tilted lattice onto a chain of interacting dipoles
(doublon-hole pairs, in a singly occupied Mott shell), and then onto a chain
of spin-1/2 particles with AF Ising interactions in longitudinal and
transverse fields. In a homogeneously tilted lattice, the 1D Bose-Hubbard
Hamiltonian reads:
\begin{align*}
    H & = -t \sum_j\left(a_j^\dagger a_{j+1} + a_j a^\dagger_{j+1}\right) \\
    & \qquad + \frac U2 \sum_j n_j \left(n_j - 1\right) - E\sum_j j \cdot n_j
\end{align*}
Here $t$ is the nearest-neighbor tunneling rate, $U$ is the onsite
interaction, $E$ is the tilt per lattice site, $a_j^\dagger$ ($a_j$) is the
creation (annihilation) operator for a particle on site $j$, and $n_j =
a^\dagger_j a_j$ is the occupation number operator on site $j$.

For a tilt near $E=U$, the onsite interaction energy cost for an atom to
tunnel onto its neighbor is almost precisely cancelled by the tilt energy. If
one starts in a Mott insulator with $M$ atoms per site, an atom can then
resonantly tunnel onto the neighboring site to produce a dipole excitation
with a pair of sites with $M+1$ and $M-1$ atoms. The resonance condition is
only met if adjacent sites contain equal numbers of atoms, so only one dipole
can be created per link and neighboring links cannot both support dipoles. We
define a (properly normalized) dipole creation operator $d_j^\dagger = \frac{a_j
a_{j+1}^\dagger}{\sqrt{M(M+1)}}$.

The Bose-Hubbard Hamiltonian above can hence be mapped onto the dipole Hamiltonian:
\[H = -\sqrt{M(M+1)} t \sum_j \left(d_j^\dagger + d_j\right) + \left(U-
E\right) \sum_j d_j^\dagger d_j\] 
subject to the constraints $d_j^\dagger d_j \leq 1$, $d_{j+1}^\dagger d_{j+1}
d_j^\dagger d_j = 0$.

The factor of $\sqrt{M(M+1)}$ arises due to bosonic enhancement.

To map from the dipole Hamiltonian to the spin-1/2 Hamiltonian, we define a
link without (with) a dipole excitation to be an up (down) spin along $\hat
z$. Then the creation/annihilation of dipoles are related to the flipping of
spins, and we can write:
\[S_z^j = \frac12 - d_j^\dagger d_j, S_x^j = \frac12\left(d_j^\dagger +
d_j\right), \text{and}\, S_y^j = \frac i2 \left(d_j^\dagger - d_j\right)\]

The constraint forbidding adjacent dipoles can be implemented by introducing a
positive energy term $Jd_{j+1}^\dagger d_{j+1} d_j^\dagger d_j$ to the
Hamiltonian, where $J$ is of order $U$. This term gives rise to
nearest-neighbor interactions and an effective longitudinal field for the
spins.

Defining $\Delta = E- U$ the Hamiltonian for the spins now reads:
\begin{align*}
    H & = J\sum_j S_z^j S_z^{j+1} - 2 \sqrt{M(M+1)} t \sum_j S_x^j \\
    & \qquad - \left(J- \Delta\right) \sum_j S_z^j \\
    & = J \sum_j \left(S_z^j S_z^{j+1} - h_x S_x^j - h_z S_z^j\right)
\end{align*}
The dimensionless fields are defined as $h_x = 2^{3/2} t/J = 2^{3/2} \tilde
t$, $h_z = \left(1 - \frac{\Delta}J\right) = 1 - \tilde \Delta$, with $M$ set
to one as in our experiment.

\subsection{Experimental Details}
Our experiments start with a single layer 2D Mott insulator of
$^{87}$Rb atoms in a $35E_r$ lattice with 680 nm spacing as described in previous work. The atoms are in the 
$|F=1, m_f=-1\rangle$ state and the initial fidelity of the Mott insulator is
0.95(2), with local fidelities as high as 0.98. A magnetic field gradient
along the $x$-direction is ramped up within 8 ms to tilt the lattice potential
by $0.7U$ per lattice site. At this point, the depth of the lattice along the
chains is ramped down to $14E_r$, while the potential transverse to the chains
is ramped up to $45E_r$, within 2 ms. At the same time, the optical potential
providing harmonic confinement is ramped down. Tunneling between chains is
negligible over the experimental timescale (see Supplementary notes for
further discussion). The gradient is then ramped adiabatically through the
transition point using a linear ramp that ends at a tilt of $1.2U$ per lattice
site, typically within 250 ms.

We can then perform either an \textit{in situ} measurement or a 1D expansion of the
chains to achieve noise correlation interferometry. In both cases, we use
fluorescence imaging after pinning the atoms in a deep lattice to obtain the
density distribution with single atom/single lattice-site resolution. Images
far on the Mott side of the transition are used to select chains of atoms
within the first shell of the insulator. The phase transition is then studied
only within these chains, with quantitative curves employing data only from
the single chain with lowest disorder.

For noise correlation measurements, the magnetic field gradient and the
lattice along the chain are switched off, while the interchain lattice and the
potential confining the atoms in the third direction remain on. After a 1D
expansion for 8 ms, the atoms are pinned for imaging. To extract information
about density wave ordering in the chains, several hundred images (250 for
paramagnetic, 500 for antiferromagnetic phase) each containing 15 chains, are
fitted to extract the atom positions, and then spatially autocorrelated and
averaged as described in Ref. \cite{foelling_strongly_correlated}.
 
\begin{figure}
    \centering
    \includegraphics[width=8.5cm]{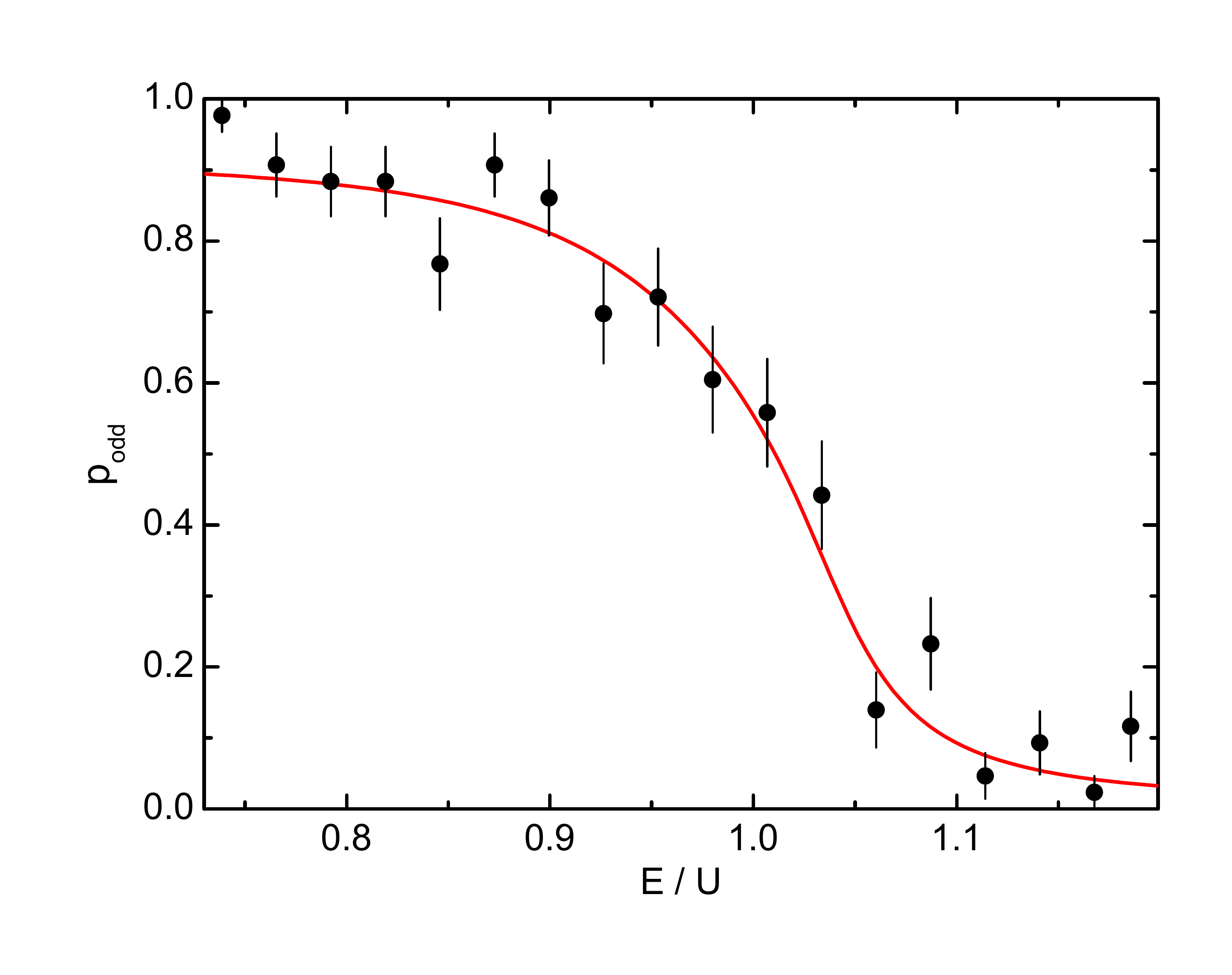}
    \caption{\label{supfig1} \textbf{Single-site transition curve.} The
    occupation probability $p_{\text{odd}}$ of a characteristic single site,
    plotted versus tilt as the system is ramped from the PM phase into the AF
    phase. The theory curve reflects a zero temperature exact diagonalization
    calculation of the ground state of a chain of six Ising spins (the shape
    of the $p_{\text{odd}}$ curve is insensitive to chain length, see Supplementary
    Fig.~\ref{supfig3}), with periodic boundary conditions. The curve has been
    offset and rescaled vertically to account for defects arising from both
    the initial MI, and heating during the ramp. The theory allows us to
    extract a lattice depth of $14(1)E_r$. We attribute the residual
    fluctuations around the expected curve to residual oscillations reflecting
    non-adiabaticity arising from that fact that the ramp was initiated too
    close to the transition. The error bars are $1\sigma$ statistical
    uncertainties.}
\end{figure}

Lattice depths are calibrated to 15\% using Kapitza-Dirac scattering, however
the width of single-site transition regions was found to be a more sensitive
probe of the longitudinal tunneling rate and hence the longitudinal lattice
depth (see Supplementary Fig.~\ref{supfig1}), and accordingly was employed
throughout this article.

\begin{figure}
    \centering
    \includegraphics[width=8.5cm]{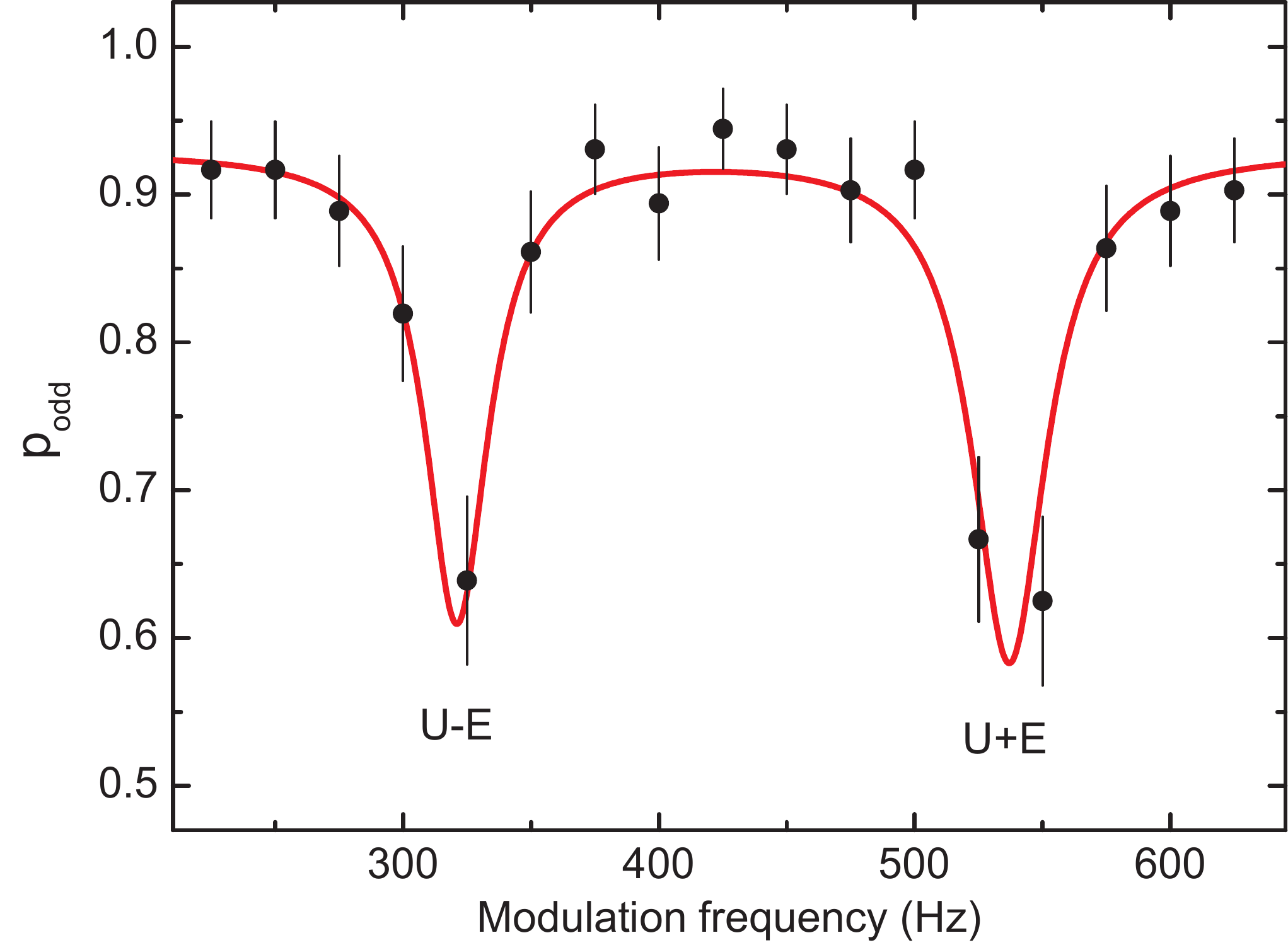}
    \caption{\label{supfig2} \textbf{Modulation spectroscopy in a tilted
    lattice.} The occupation probability averaged over the six-site
    near-homogeneous region described in the main text, plotted versus the
    modulation frequency, for $16E_r$ longitudinal lattice modulated by
    $\pm23\%$, corresponding to a Bose-enhanced resonant tunneling
    rate\cite{esslinger_nearest_neighbor} of $2\pi \times 4 \, \text{Hz}$. Because
    the experiment is performed in a lattice tilted by $E$ per site, the peak at
    zero tilt which appears at the interaction energy $U$ is split out into two
    peaks, one corresponding to an atom tunneling up the tilt at an energy
    cost of $U+E$, and one to tunneling down the tilt a cost of $U-E$. Fitting
    these peaks allows us to extract both $U$, and $E$. The peak width arises
    from a combination of power broadening (approximately $2\pi \times 14
    \, \text{Hz}$, complicated by Rabi flopping), and the residual lattice
    disorder discussed in the main text.}
\end{figure}

The magnetic field gradient is calibrated using lattice modulation
spectroscopy. In the presence of a potential gradient $E$ per lattice site,
modulation of the lattice depth along the chains causes resonant excitation at
two frequencies, $U+E$ and $U-E$ corresponding to an atom in the Mott
insulator moving up or down gradient. We detect these excitations as a
reduction in the value of $p_{\text{odd}}$ using in-situ imaging (see
Supplementary Fig.~\ref{supfig2}). Using the mean of the two resonances, we
obtain the interaction energy $U=430(20)\,\text{Hz}$ at $16E_r$ longitudinal
lattice, $45E_r$ transverse lattice (corresponding to $U=413(19)\,\text{Hz}$
at $14E_r$ longitudinal lattice, where the experiment operates, which agrees
with a band-structure calculation of 401(25) Hz). The separation between the
resonances as a function of applied gradient is used to calibrate $E$. At zero
applied magnetic field gradient, we find the stray gradients to be less than
$0.02U$.

\subsection{Local and Long-range Observables}
\begin{figure}
    \centering
    \includegraphics[width=8.5cm]{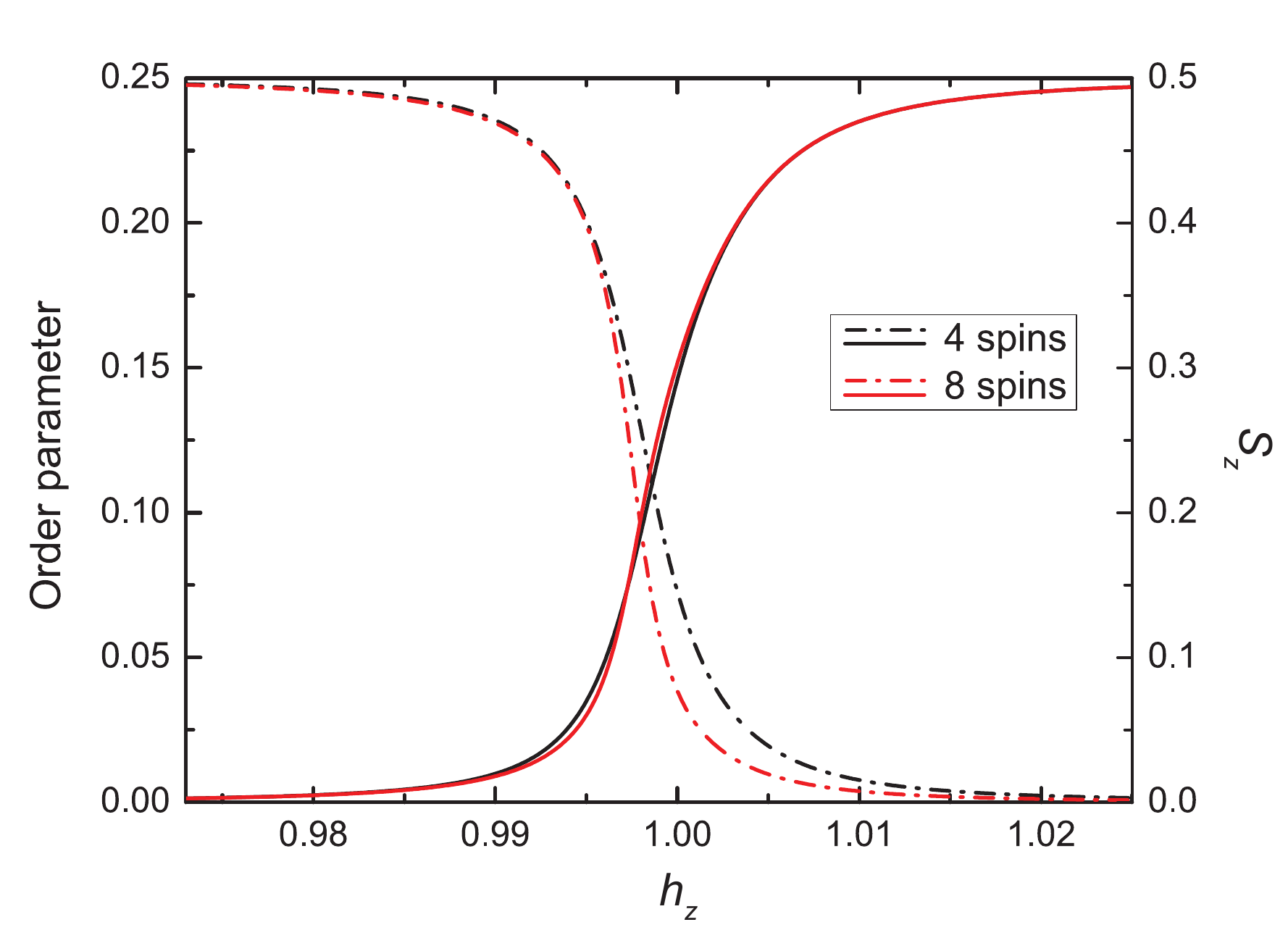}
    \caption{\label{supfig3} \textbf{Comparing $\bm{S_z}$ to the order
    parameter.} An exact diagonalization calculation (for $h_x=0.004$) of the
    ground state of a 1D chain of four (black) and eight (red) Ising spins with
    nearest-neighbor interactions, revealing that $S_z$ (solid) is not sensitive
    to atom number, while the order parameter (dash-dotted) is. It is
    anticipated that the order parameter will exhibit a cusp in the large-system
    limit, though the exponential scaling with atom number precludes
    simulating substantially larger systems on a classical computer.}
\end{figure}

In-situ detection gives the atom number modulo 2 due to light assisted
collisions of atoms on each lattice site during imaging. By averaging the
occupation of a site over multiple images, we obtain the probability of an odd
occupation on the $j$th lattice site ($p^j_{\text{odd}}$), which corresponds
to the probability of having a single atom on a site within the subspace of
our model. This is related to the spin observables in the effective model by
$\langle S_z^{j-1} S_z^j\rangle = \frac12 \left(p_{\text{odd}}^j - \frac12
\right)$. We average over all the atoms in the chain to obtain $p_{\text{odd}}
= \overline{p^j_{\text{odd}}}$, which in combination with the constraint that
neighboring down spins are not allowed permits us to relate the chain averaged
mean $z$-projection of spin to $p_{\text{odd}}$ according to:
$\overline{\langle S_z^j \rangle} = \frac{p_{\text{odd}}}2$. This quantity varies across the transition and depends only weakly on the chain length. On the other hand, the order parameter for the transition $O = \biggr\langle\biggr( \frac1N \sum_j (-1)^j S_z^j
\biggr)^2\biggr\rangle$
depends on the chain length (Supplementary Fig.~\ref{supfig3}) and is
non-analytic across the transition for a thermodynamic system.
The amplitude of the noise correlation signal at separation $d$ for a chain
consisting of a large number of atoms is $C(d) = 1 + \frac1{N^2} \biggr| \sum_j e^{-i \frac{madj}{\hbar t}}
n_j\biggr|^2$ where $N$ is the number of lattice sites, $j$ is the lattice
site index, $a$ is the lattice spacing, $m$ is the mass of the atom, $t$ is
the expansion time, $n_j$ is the occupation of the $j$th site. It can be shown
that the correlation signal at $d = \frac{\pi \hbar t}{ma}$ is related to the
order parameter $O = C\left(\frac{\pi \hbar t}{ma}\right) - 1$.
\section{Supplementary Information}
\subsection{Supplementary Discussion of Dark Domain Length Analysis}
The magnetic interactions should produce even-length domains of dark sites,
corresponding to AF spin domains. To quantify the length of these AF domains
we study the length of the measurable dark domains, defining a ``dark domain''
as a contiguous string of dark sites that is bounded either by a site with an
atom or an edge of the region of interest. We then calculate the mean
length-weighted dark chain length from this data.

\begin{table*}[!h]
    \centering
    \begin{ruledtabular}
    \begin{tabular}{....} 
        \multicolumn{1}{c}{Mott Fidelity $p_{\text{odd}}$} & 
        \multicolumn{1}{c}{Entropy per Particle ($S/Nk_B$)} &
        \multicolumn{1}{m{4cm}}{Length-Weighted Mean AF Domain Size (Thermalized)} & 
        \multicolumn{1}{m{4cm}}{Length-Weighted Mean Uninterrupted Chain Length (Unthermalized)} \\
        \hline
        0.95 &  0.23 & 22 & 39 \\
        0.975 & 0.13 & 52 & 79 \\
        0.99 &  0.063 & 144 & 199 \\
    \end{tabular}
    \end{ruledtabular}
    \caption{\label{suptable} For various Mott insulator fidelities, the
    corresponding configurational entropy per particle is computed. These
    entropies are comparable to, or well below, the critical entropy for
    quantum magnetism\cite{capogrosso_critical_entropy} $S/Nk_B\sim0.25-0.5$.
    If the spin degrees of freedom thermalize efficiently with the Mott
    degrees of freedom, the spin entropy will then be equal to the Mott
    entropy. The corresponding mean AF domain size is then computed for each
    Mott entropy. In the absence of thermalization, the Mott defects break the
    spin chain into disconnected subsystems, whose mean size is computed in
    the fourth column, and is comparable to the mean chain length in the
    presence of thermalization.}
\end{table*}

Defects in the initial Mott insulator (MI) reduce the effective system size.
Their appearance can produce an overestimate of the dark chain length by
either connecting two dark chains, or appearing on the end of a dark chain.
The initial MI defect probability is typically 4\% per site over an entire
$N=1$ shell, after correcting for losses during imaging.

Losses and higher order tunneling processes during the ramp can have similar
consequences for the observed dark domain length, and can also suppress the
observed dark domain length by perturbing atoms near the end of the ramp once
the AF has already formed. The rate of such processes can be estimated from
the MI $1/e$ squeezing lifetime in the tilted lattice, measured to be 3.3
seconds. To perform this measurement we first ramp to a tilt of 300 Hz/lattice
site and tune the lattice depths to $45E_r$ and $14E_r$ for transverse and
longitudinal lattices, respectively. We then hold for a variable time, and
measure the observed \textit{in situ} atom number. The lifetime is dominated
by higher-order tunneling processes. Parametric heating and inelastic
scattering become the dominant loss channel in deeper lattices once tunneling
is inhibited.

A worst-case estimate for the impact of missing atoms can be reached from the
fraction of the time that the system is missing no atoms at the end of the
ramp. The six-site chain analyzed in the main text is initially fully occupied
79\% of the time. During the time it takes the dark-domain length to grow to 4
lattice sites (60 ms), the aforementioned effects only reduce this number to
73\%.

\subsection{Supplementary Discussion of Entropy and Thermalization}
Supplementary Table 1 shows the entropy per particle ($S/Nk_B$) for several
different Mott insulator fidelities ($p_{\text{odd}}$), assuming a chemical
potential $\mu = U/2$, as well as the mean length-weighted AF domain size $D$
in an infinite 1D magnetic system with the same entropy per particle. Here $D
=2(2-\epsilon)/\epsilon$, where the spin-dislocation probability in the AF
$\epsilon$ is defined by $S/Nk_B \approx (\epsilon/2)\left[1+
\log(1/\epsilon)\right]$. Spin defects are ignored as they are both
dynamically and thermodynamically unlikely. The entropy per particle can be
related to the Mott insulator fidelity by\cite{esslinger_fermionic_MI}:
$S/Nk_B=\log\left[\frac2{1-p_{\text{odd}}}\right] - p_{\text{odd}}
\log\left[\frac{2p_{\text{odd}}}{1-p_{\text{odd}}}\right]$. If such
thermalization took place in our finite length chain of six sites (with initial
fidelity 97.5\%), the mean domain size would be limited by the system size to
5.3 sites.

Experimentally, we find most Mott defects to be unbound doublons and holes,
which do not directly map to excitations in the spin model. The large energy
gap present in our tilted lattice, combined with conservation of particle
number, make it difficult for these Mott defects to thermalize with spin
degrees of freedom. Such thermalization would require, for example, migration
of a doublon to a hole, or decay via a very high order
process\cite{capogrosso_critical_entropy} into several spin defects- quite
unlikely within the experimental timescale. Consequently, these nearly static
defects act as fixed boundary conditions that limit the effective length of
the simulated spin chains. Supplementary Table one also provides the expected
uninterrupted chain length, computed as $L_{\text{sys}} =
(1+p_{\text{odd}})/(1-p_{\text{odd}})$.

\subsection{Supplementary Discussion of Higher Order Effects}
\subsubsection{Interchain Tunneling}
Tunneling between chains is excluded from the spin-mapping described in the
main text, though under certain conditions it produces exotic transverse
superfluidity, as described in Ref. \cite{sachdev_MI_electric_fields}. For our
purposes, these tunneling processes serve only to take the system out of the
Hilbert space described by the spin model. Most of our experiments were
performed at a transverse lattice depth of $45E_r$, corresponding to an
interchain tunneling rate of $t_{\text{transverse}} = 2\pi \times 0.07 \,
\text{Hz}$. This tunneling rate is basically negligible on our experiment
timescale of 250 ms. The noise correlation data, as well as the shell pictures
and reversibility curve in Fig.~\ref{fig2} of the main text, were taken at
$35E_r$ transverse lattice depth. At this depth the transverse tunneling rate
is $t_{\text{transverse}} = 2\pi \times 0.27 \, \text{Hz}$, which is small
compared to our lattice inhomogeneities, and so results in highly-suppressed,
off-resonant Rabi-flopping. In practice, increasing the transverse lattice
from $35E_r$ to $45E_r$ results in a modest $\sim5\%$ improvement in the
quality of the Mott insulator after transitioning to the antiferromagnetic
state and back.

\subsubsection{Second Order Tunneling}
In addition to nearest-neighbor tunneling which creates doublon-hole pairs,
and proceeds at a rate $\sqrt2 t$ when the tilt $E=U$, there remain
second-order tunneling processes which create triplons at a rate
$t_{\text{S.O.}} \sim \frac{\sqrt 3 t^2}{U}$. For our longitudinal lattice
depth of $14E_r$, and interaction energy $U = 2\pi \times 416 \, \text{Hz}$,
we find $t_{\text{S.O.}} \sim 2\pi \times 0.4 \, \text{Hz}$.

Because our system is continuously tilted, all such transitions will be tuned
through resonance. For our typical experiment, $R_{\text{ramp}} \approx
\frac{\frac12 U}{250 \, \text{ms}} \approx 2\pi \times 840 \, \text{Hz}^2$, so
the Landau-Zener adiabatic transition probability to the triplon state
$P_{\text{triplon}} = 1 - \exp \left[-2\pi \times
\frac{t_{\text{S.O.}}^2}{R_{\text{ramp}}}\right] \sim 1\%$. In future experiments
with slower ramps, both this effect and the closely related second-order
Stark-shift will become more of a concern. These can be further suppressed
relative to the desired dynamics by increasing the longitudinal lattice depth,
at the expense of slower many-body dynamics. It bears mentioning that for our
experimental parameters the triplon state should experience an additional
energy shift calculated to be 22 Hz, due to multi-orbital
interactions\cite{will,johnson}

\subsubsection{Impact of physics beyond the Hubbard Model}
For a $14E_r$ lattice, the next-nearest neighbor tunneling rate is suppressed
relative to that of the nearest neighbor\cite{jaksch} by a factor of
$\sim300$, making the total rate $t_{\text{NextNeighbor}} = 2\pi \times 0.04
\, \text{Hz}$, which is negligible on present experiment timescales. The
longitudinal nearest-neighbor interaction shift for one atom per lattice site
is $\sim 10^{-3} \, \text{Hz}$, and interaction driven
tunneling\cite{foelling_second_order} occurs with a rate of $2\pi \times 0.3
\, \text{Hz}$.

\end{document}